\documentstyle[psfig]{mn}
\newif\ifAMStwofonts
\AMStwofontstrue

\title{The radio source counts at 15~GHz and their implications for cm-wave CMB imaging}
\author[Angela C. Taylor et al.]
       {Angela C. Taylor, Keith Grainge, Michael E. Jones, G. G. Pooley,
Richard Saunders,
\newauthor
and E. M. Waldram \\Astrophysics Group, Cavendish Laboratory,
Madingley Road, Cambridge, CB3 OHE, UK}
\begin{document}

\maketitle

\label{firstpage}

\begin{abstract}
We present the preliminary results of a new survey of radio sources
using the Ryle Telescope at 15.2~GHz. This is the highest frequency at
which a survey has been done that is relevant to the issue of radio
source contamination in CMB experiments.  The differential source
count of the 66 sources found in 63~deg$^{2}$ is 80$(S/\rm
Jy)^{-2.0}Jy^{-1}sr^{-1}$, from $\approx$~20~mJy to $\approx$~500~mJy.  Extrapolating this to 34~GHz (where
many cm-wave CMB experiments operate) gives an estimated temperature
contribution of sources ${\Delta}T_{\rm conf}$~=~9$\mu$K in a CMB image, with a
beam corresponding to multipole $l\approx$~500.  A means of source
subtraction is evidently necessary, otherwise the signal-to-noise ratio in
CMB images will be limited to 4 or 5, becoming worse at higher
resolution.  We compare the population of sources observed in this
new survey to that predicted by extrapolation from lower frequency
surveys, finding that source fluxes, and indeed the existence of many
sources, cannot be determined by extrapolation.
 
\end{abstract}


\section{Introduction}
A fundamental limitation to all measurements of primordial Cosmic
Microwave Background (CMB) anisotropies is that set by astronomical
foregrounds, both from the Galaxy and from extragalactic sources.  The
degree to which extragalactic radio sources contaminate is dependent on both the resolution and observing frequency of a particular
experiment.  For example, owing to relatively low collecting area and angular
resolution, the $COBE/DMR$ results \cite{cobe} are
essentially unaffected by extragalactic sources.  Yet as the new
generation of CMB experiments come on-line and strive to reach higher
resolution and sensitivity, foreground sources are becoming an increasingly
important consideration.

This can be seen simply by considering the Rayleigh--Jeans approximation for the flux density of an unresolved source as a
function of the angular resolution, $\Omega^{-1/2}$, of an experiment: 
\begin{equation}
 S=\frac{2k_{\rm B}T\Omega}{{\lambda}^2},
\end{equation}
where $S$ is the source flux density, $k_{\rm B}$ is the Boltzmann constant,
$\Omega$ is the beam solid angle, $\lambda$ the
observing wavelength and $T$ is the resulting antenna temperature.
Expressing this in terms of multipole index $l$ (where $l\simeq
{2\pi}/{\theta}$ for high $l$)
we see that the temperature contribution of a source increases as $l^{2}$:
\begin{equation}
 T=\frac{Sl^{2}\lambda^{2}}{8k_{\rm B}\pi^{2}}.
\end{equation}
Since sources are distributed randomly on the sky, the power spectrum
of point sources also increases as $l^{2}$.  
The correct removal of point sources therefore becomes more important as the CMB power spectrum is investigated on smaller
angular scales, a regime in which there is even greater interest since the
recent Boomerang \cite{boomerang} and MAXIMA \cite{maxima} results.

The removal of foreground sources, however, is not trivial. They are known to have a wide range of spectral indices. At low
frequencies, their spectral indices $\alpha$ (with $S{\propto}~{\nu}^{-\alpha}$) are usually steep ($\alpha \ge 0.5$), but some are
flat ($\alpha$~=~0) or may even be rising. Very occasionally, a
spectrum may rise so steeply that it mimics that of the CMB with
$\alpha$~=~$-$2 \cite{edge}.

The secure method of determining the point sources present in a CMB
observation is to map at high resolution.  CMB experiments, however,
operate at frequencies ($\ge$ 30~GHz) where the population of point
sources is not well characterised. Indeed, until the present results,
there existed no suitable high-frequency survey above 4.8~GHz
\cite{gb6}.  Windhorst et al. \shortcite{wind} provide counts at 8.4~GHz of the
sub-mJy source population, but these are too faint to affect
primordial CMB
studies.  They also plot the 10.0-GHz data of Aizu et al. \shortcite{aizu},
spanning 30--3000~mJy. However, as Aizu et al. state, their 10.0-GHz
observations only include sources selected at 5~GHz and do not
represent an independent survey.  Theoretical predictions of the
extragalactic source counts at a wide range of frequencies have also been
made \cite{toff}, but observations at relevant frequencies
are still needed to confirm the results.  Furthermore, although source
counts enable one to predict the point source contribution to a
measured CMB power spectrum, this is helpful only if the source
contribution is small compared with the CMB power.  In general,
finding the sources with high-resolution imaging is necessary.

In this paper we present the preliminary results of a source survey at
15~GHz that is part of the source-subtraction
strategy for a new CMB interferometer, the Very Small Array (VSA,
see Jones \& Scott 1998).  This is the highest frequency at which a survey has been
undertaken that is relevant to the issue of radio-source
contamination in cm-wavelength CMB experiments.  The spectra and number count of the
sources are discussed, as are the implications for the
removal of extragalactic sources from CMB measurements.  We stress
that the temperature confusion estimates presented here apply to
images of the CMB; we shall present estimates of the effects of sources on the
CMB power spectrum in a subsequent paper. 

\section[]{Source Subtraction for the VSA\\ and the first 15-GHz survey}
\label{sec:source-subtr-vsa}
The VSA is a close-packed, 14-element interferometer for sensitive
measurement of CMB anisotropies.  In its compact configuration the VSA
will measure the CMB power spectrum in the $l$-range 120--700, whilst
in its extended array it will reach $l$~=~1800.  The telescope is
currently operating at 34.1~GHz and we expect extragalactic radio
sources to be a major contaminant in VSA observations.  We
have developed a strategy to deal with this problem in two steps.

First, we use low-frequency surveys, NVSS at 1.4~GHz \cite{nvss} and Green
Bank at 4.8~GHz \cite{gb6}, to select CMB fields in which there are predicted to be no
sources brighter than 500~mJy at 34~GHz.  Predictions are made by
extrapolating the fluxes of every source in the 4.8-GHz catalogue to
34~GHz on the basis of its spectrum between 1.4 and 4.8~GHz.  The VSA
fields are also selected to be as free as possible of other Galactic and
extragalactic foregrounds.  In particular they have been chosen for high
galactic latitude, low galactic synchrotron and free-free emission,
low dust and putative dust, and a lack of nearby large-scale
structure.

Having selected suitable fields, and prior to VSA observation, we locate potentially contaminating
radio sources by mapping each VSA field at a much higher
resolution than that of the VSA.  This is achieved using the Ryle Telescope (RT) in Cambridge, operating at 15.2~GHz in a
raster-scanning mode \cite{ryle}.  We reach an rms noise level
of $\sigma$~=~4~mJy in order to identify all sources above 20~mJy at
15.2~GHz.  Simultaneously with VSA observing, we then monitor
the flux of each source using a separate single-baseline
interferometer operating at the same frequency as the VSA.  This
interferometer uses two 3.7-m antennas located next to the VSA, using
identical receivers and correlator to the main array.  The
monitoring is done simultaneously with VSA
observations so that variable sources can be subtracted accurately.

\section{Results and implications of the\\ 15-GHz survey}
\begin{figure}
\psfig{file=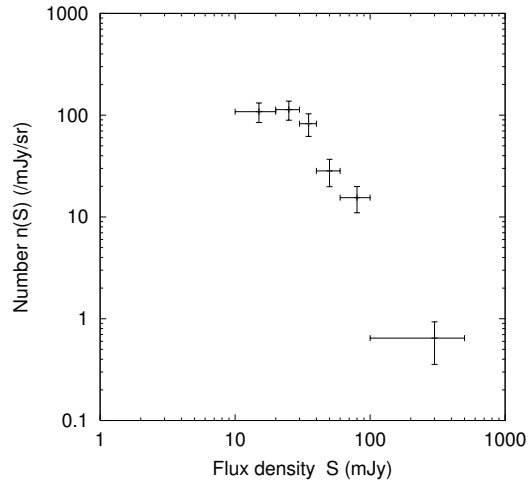,height=6.5cm,angle=-90}
\caption{The differential source count of sources observed in VSA1 at
15.2~GHz.  The fall-off at low S is ascribed to the lack of
completeness.}
\end{figure}
The first VSA field, VSA1, covering an area of 63~deg$^{2}$ and centred
at 00$^{h}$20$^{m}$+~30$^\circ$ (B1950), contains 66 sources with $S~\ga$~20~mJy.
The differential source count for this field is shown in Figure 1 and,
between 20 and 500~mJy, is well
fitted by 
\begin{equation} 
n(S)=8\left( \frac{S}{100~{\rm mJy}}\right)^{-2.0}\rm mJy^{-1}\rm sr^{-1}\equiv~80\left(\frac{S}{\rm Jy} \right)^{-2.0}~\rm Jy^{-1}\rm sr^{-1}
\end{equation}
The spectral index distribution between 1.4~GHz and 15~GHz for all
the sources found in our survey is shown in Figure 2.  A significant
fraction of these sources were not predicted to be seen at all; they
have components with flat or rising spectra that only surface above
4.8~GHz. Note that in order to assign a spectral index to all the
sources, we have had to use ${\alpha}_{1.4}^{15}$ rather than ${\alpha}_{4.8}^{15}$ since the
Green Bank 4.8-GHz survey is far less deep than NVSS.  A
consequence of this is that ${\alpha}_{4.8}^{15}$ is probably even
steeper than ${\alpha}_{1.4}^{15}$ for a few of the sources.
\begin{figure}
\psfig{file=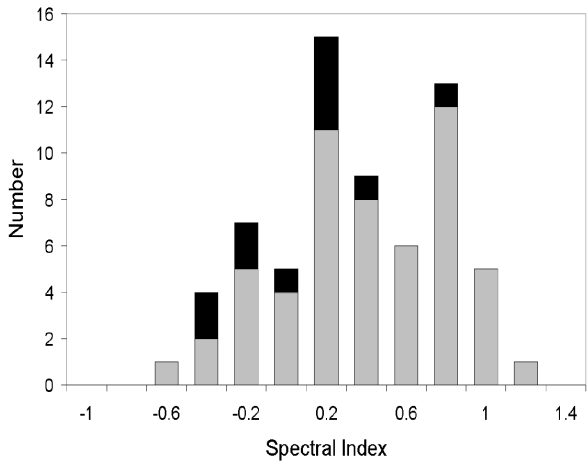,height=6.5cm,width=8cm}
\caption{Spectral index distribution, $\alpha_{1.4}^{15}$, of
sources found at 15.2~GHz with the Ryle Telescope. Grey bars represent
observed sources predicted
by extrapolation of spectra using the 1.4-GHz NVSS and
4.8-GHz Green Bank surveys.  The black bars represent those sources observed but {$not$} predicted.}
\end{figure}

To estimate the effect of these sources on CMB images, we follow
Scheuer \shortcite{scheuer}:
\begin{equation} 
{\sigma}_{\rm conf}^{2}~=\Omega\int_{0}^{S_{\rm lim}}n(S) S^{2} dS,
\end{equation}
where $\sigma_{\rm conf}$ is the confusion noise in flux density from unsubtracted sources
and $S_{\rm lim}$ is the flux density above which all sources have been
subtracted.  For $l$~=500 (typical of the VSA compact array and of
other cm-wave experiments) and assuming a Gaussian distribution of
baselines, $\Omega~=(2\pi/500)^2$ giving a 43-arcmin beam.  If $no$
source subtraction is attempted
(i.e. $S_{\rm lim}$ $\approx$ 500~mJy, and we emphasize here that this VSA field has
been pre-selected to avoid bright sources), then equation (4) predicts
$\sigma_{\rm conf}$~=~79~mJy at 15~GHz.

We now consider what $\sigma_{\rm conf}$ might be at the VSA
observing frequency of 34~GHz.  Our educated guess of the mean
spectral index between 15 and 34~GHz, given the change in
source population from 4.8 to 15~GHz that we have found at these flux
levels, is that $\alpha_{15}^{34}$~=~0.5.  This is less steep than the typical
$\alpha$ at low frequencies but reflects the increasing fraction of
flat- and rising-spectrum sources at high frequency.  Assuming this
value of $\alpha$, we estimate
$\sigma_{\rm conf}$ at 34~GHz to be 52~mJy in a 43-arcmin beam
corresponding to $l\approx$~500.  Equation
(1) then implies that the resulting temperature from unsubtracted
sources, ${\Delta}T_{\rm conf}$, is 9.2~$\mu$K. This would limit the
signal-to-noise ratio in CMB images to 4 or 5, and the problem
becomes even worse at higher resolution.  Source subtraction is
evidently necessary.

By surveying for sources one can obviously decrease $S_{\rm lim}$.  If a
survey reaches $S_{\rm lim}$~=~50~mJy at 15~GHz, then Equations (1) and
(4) and the assumed $\alpha_{15}^{34}$~=~0.5 give
${\Delta}T_{\rm conf}~$=~2.9~$\mu$K at 34~GHz in a beam corresponding to
$l~\approx$~500.  The fact that the RT survey reaches $\approx$
20~mJy means that it will find all sources which have
$\alpha_{15}^{34}$ rising as steeply as $-$1.  This is likely to include
the majority of the source population.  Sources with $\alpha$
approaching $-$2, however, are known to exist (see e.g. Edge et
al. 1998). An upper limit to the contribution of such sources can be
found as follows.  A source with $\alpha_{15}^{34}$~=~$-$2 and
$S_{15}$~=~20~mJy will have $S_{34}$~=~103~mJy.  If we assume that $all$
sources have $\alpha_{15}^{34}$~=~$-$2 and that $n(S)$ is given by
Equation (3), then the RT survey effectively has $S_{\rm lim}$~=~103~mJy at
34~GHz.  This results in ${\Delta}T_{\rm conf}$~=~6.4~$\mu$K in a beam
corresponding to $l\approx$~500. The population will of course not be this
extreme, and ${\Delta}T_{\rm conf}$ for VSA compact array maps
is likely to be close to 3~$\mu$K.
\begin{figure}
\psfig{file=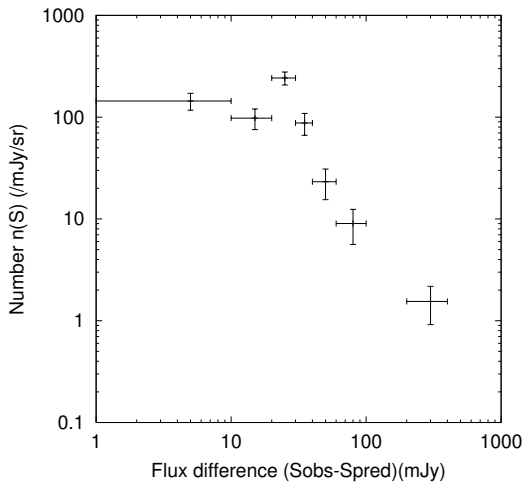,height=6.5cm,angle=-90}
\caption{The differential count of residual flux remaining after
source subtraction has been done using predicted source fluxes.}
\end{figure}

Finally, we consider the effect of source subtraction using only predicted
source fluxes and positions extrapolated from low frequency
surveys. Using this approach we expect to find
122 sources with a flux density greater than 20~mJy within our survey
region of 63~deg$^{2}$.  We find, however, that only 55 of the predicted sources are actually found, the
rest having spectra that fall off enough between 4.8 and 15~GHz to
drop them below our flux limit.  Furthermore, 11 sources were found in
our survey which were not predicted by extrapolation.  Five of the
unpredicted sources do not even appear in the 4.8-GHz Green Bank
catalogue. To find the residual source flux that would be left on an
image if only predicted source fluxes had been subtracted, we
calculate the modulus of the difference between predicted and observed flux
for each source.  Figure 3 shows the differential count of this residual flux;
note that the remaining source contamination is somewhat worse than if no
source subtraction had been attempted.  Specifically, we find that substantial errors in individual source fluxes
are made, both in over-predicting the fluxes of a large fraction of
sources with down-curving spectra, and non-prediction of a significant
fraction of sources with rising spectra. This result, however, is not unexpected.  As we move to higher
frequency regimes, we expect many more sources to be
synchrotron-self-absorbed (SSA), exhibiting intrinsically rising spectra and
increased variability.  Since the surveys we used to predict the
source population at 15~GHz are a decade lower in frequency, one
would therefore expect to underestimate the number of SSA
sources, leading to a non-prediction of many sources with rising
spectra.  Furthermore, the surveys used were undertaken at
different epochs and hence no allowance can be made for variable sources.

\section{Conclusions}
\begin{enumerate}
\renewcommand{\theenumi}{(\arabic{enumi})}
\item We have used the Ryle Telescope at 15.2~GHz to survey 63~deg$^{2}$ of
sky chosen, on the basis of
low--frequency information, to contain no source with flux density
$>$~500~mJy.  The differential source count of the 66 sources found is 
80$(S/\rm Jy)^{-2.0}Jy^{-1}sr^{-1}$ from $\approx~$20~mJy to $\approx$~500~mJy.
\item Extrapolating the source count to 34~GHz, near to which current cm-wave CMB experiments
operate, the temperature contribution of sources to a CMB image with a
43-arcmin beam (corresponding to $l\approx$~500) is estimated to be
${\Delta}T_{\rm conf}$~=~9$\mu$K.  Hence, the signal-to-noise ratio in CMB images
will be limited to 4 or 5, and even more at higher resolution,
if foreground sources are not removed.
\item In the case of the VSA compact array, which will measure the CMB power
spectrum in the $l$-range 120--700, the RT source surveying should
reduce ${\Delta}T_{\rm conf}$ to 3~$\mu$K in a 43-arcmin beam at 34~GHz.
\item Our results show that the existence and flux density of individual sources
cannot be predicted accurately by extrapolation from low-frequency
surveys.  We find that trying to predict source fluxes at 15~GHz by
extrapolating their 1.4 and 4.8-GHz fluxes gives a greater error,
i.e. a greater ${\Delta}T_{\rm conf}$, than attempting no source subraction
at all.
\end{enumerate}
\section*{Acknowledgments}
We thank the staff of the Cavendish Astrophysics group for operation of the Ryle Telescope, which is funded by PPARC.  ACT
acknowledges a PPARC studentship.

\label{lastpage}
\end{document}